\documentclass[preprint,aps]{revtex4}
\usepackage{graphicx,pdflscape,epsf,epsfig,amsmath}
\usepackage{alltt,dsfont,amsmath,amssymb,bm}

\newcommand{\G}{{\cal{G}}}
\newcommand{\GH}{{\bf g}}
\renewcommand{\l}{{ (local)}}
\newcommand{\tJ}{\ $t$-$J$ \ }
\newcommand{\Q}{{\cal Q}}
\newcommand{\nn}{\nonumber}
\newcommand{\chem}{{\bm \mu}}

\newcommand{\beq}{\begin{equation}}
\newcommand{\eeq}{\end{equation}}
\newcommand{\barray}{\begin{eqnarray}}
\newcommand{\earray}{\end{eqnarray}}
\newcommand{\disp}[1]{Eq.~(\ref{#1})}
\newcommand{\refdisp}[1]{Ref.~(\onlinecite{#1})}
\newcommand{\figdisp}[1]{Fig.~(\ref{#1})}

\begin{document}
\title{ Dynamical Particle Hole Asymmetry in  Cuprate Superconductors}
\author{ B. Sriram Shastry}
\affiliation{ Physics Department, University of California, Santa Cruz, CA 95064, USA}
\date{May 31, 2012}

\begin{abstract}

Motivated by the form of recent theoretical results,    a quantitative test for an important  dynamical particle hole asymmetry of the electron spectral function at low energies and long wavelengths is proposed.  The  test requires  the decomposition of the   angle resolved photo emission   intensity,  after a specific    Fermi symmetrization,  into  odd and even parts to obtain  their ratio ${\cal R}$.  A  large  magnitude ${\cal R}$ is implied  in recent  theoretical fits at optimal doping  around  the chemical potential, and
I  propose  that this large asymmetry  needs to be checked more directly and thoroughly.   This processing requires  a slightly higher precision determination of the Fermi momentum relative to  current availability.  
\end{abstract}

\pacs{71.10.Ay, 74.25.Jb, 74.72.Gh, 79.60.-i,71.10.Fd, }

\maketitle

\maketitle

{\bf 1. Introduction:}   
The search for a microscopic theory of the normal state of the cuprates is one of the main themes in condensed matter physics for the last two decades.
The recent suggestions of   describing     the normal state   in terms of theories with a  quantum critical point \cite{sachdev2} have  also created wide  interest in other branches of physics such as string theory and quantum gravity \cite{lee}. An  initial   theoretical objective     is   the  derivation of the  normal state  low energy long wavelength  single electron spectral function $\rho_{\G}(\vec{k}, \omega)$ (or equivalently  $A(\vec{k}, \omega)$), encoding the complete set of symmetries.

In this paper I discuss the behavior of  $\rho_{\G}(\vec{k}, \omega)$ under a   dynamical   particle hole  transformation simultaneously inverting  the wave vector  and energy relative to the chemical potential $\chem$ as:
\beq
(\vec{\hat{k}}, \omega) \to - (\vec{\hat{k}}, \omega), \;\;\; \mbox{with} \   \vec{\hat{k}} = \vec{k}- \vec{k}_F  . \label{dpht} 
\eeq
 Invariance under this transformation  has  often been invoked in analyzing  Angle Resolved Photoemission (ARPES) data\cite{mohit2}. It is an {\em emergent symmetry} of the Fermi liquid in the sense of \refdisp{schmalian},   arising when correction terms of $O(\omega/\varepsilon_F)^3$ are neglected\cite{hodges}. Fermi liquids without disorder at intermediate coupling are   invariant \cite{fndisorder} under  \disp{dpht},  as are most  other contemporary theories of  cuprates    that I am aware of.

On the other hand two recent theories, the  {\em extremely  correlated Fermi liquid} theory  (ECFL)     proposed  by the author \refdisp{ecfl},  and the {\em hidden Fermi liquid} theory due to Casey and Anderson  (CA) \refdisp{casey},   yield a spectral function that   lacks   invariance under \disp{dpht}. In \refdisp{gweon} a comparison between the ECFL spectral function  and a large set of  data at optimal doping shows excellent agreement and provide a useful parametrization of the data. 
 To  quantify the asymmetry: for optimally doped cuprates,  in an  energy range of  $ \pm 25$meV around  $\chem$, the theories and the fits of \refdisp{gweon} (extrapolated to lower $\omega$) yield an 
 {\em  asymmetry ratio } ${\cal R}$  (defined below \disp{rr}) between $\sim 7\%$  to $10\% $. Since a large asymmetry  makes a decisive ruling on the  allowed theories, we propose     the direct experimental measurement of this  effect and indicate  a procedure for the same.

 I first discuss a Fermi symmetrization procedure quite distinct  from the symmetrization in \cite{mohit2,symmetrization}.  I construct an object ${\cal S}_{\G}(\vec{k}, \omega)$ (\disp{symm}) from the observed ARPES intensity and find expressions for this in the Fermi liquid and  the ECFL model. I further  show how the momentum dependence of the  dipole transition probability  and the Fermi liquid parameter $Z_k$
 can be absorbed into the  constants.
  
 The  ${\cal S}_{\G}(\vec{k}, \omega)$ function is detailed  for a  simplified version of ECFL, 
providing  an idealized picture  of the predicted asymmetry effect in cuprates.
 I further    discuss  a related  asymmetry of the  tunneling conductance in the normal state, and also the expected {\em angle integrated} spectrum.  Within the  simplified ECFL model,   where the quasiparticle peaks are sharp over a large fraction of the zone, these  exhibit  unusual and possibly measurable features.
 
 {\bf 2. Fermi symmetrization}
 Our  first  goal  is to  formulate  a procedure for  isolating  terms  in the spectral function near the Fermi energy that are  {\em linear in wave vector and frequency} $\sim \xi_k - \omega$  (with $\xi_k = \vec{\hat{k}}. \vec{v}_{\vec{k}_F}$) found in the recent work \refdisp{ecfl}.  The  ARPES intensity is given in terms of the spectral function within the sudden approximation by the expression $
I(\vec{k}, \omega)= M(\vec{k}) \ f_\omega \ \rho_{\G}({\vec{k}}, \omega)$,
where $  M(\vec{k}) $ is the dipole  transition probability which  is expected to be a smooth function of $\vec{k}$ and   independent of   $\omega$.  It also contains  the Fermi function for occupied states $f_\omega=\{ 1+ \exp{(\beta \omega)} \}^{-1}$, a  non symmetric function of $\omega$. Therefore  we  first formulate  a {\em Fermi symmetrized object}:
\beq
{\cal S}_{\G}(\vec{k}, \omega) \equiv f_\omega  \overline{f}_\omega \rho_{\G}(\vec{k} ,\omega)  = \frac{1}{M(\vec{k})} \overline{f}_\omega  I(\vec{k}, \omega), \;\; \label{symm}
\eeq
where $\overline{f}_\omega=1-f_\omega=f_{- \omega}$.
 We may now decompose ${\cal S}_{\G}(\vec{k}, \omega)$ under \disp{dpht} into its  antisymmetric ${\cal S}_\G^{a-s}(\vec{k}_F| \vec{\hat{k}}, \omega) $ and symmetric  $ {\cal S}_\G^{s}(\vec{k}_F |\vec{\hat{k}}, \omega)$ combinations  respectively
$
 \frac{1}{2}  \ \left[ {\cal S}_\G(\vec{k}_F+ \vec{\hat{k}}, \omega)  \mp {\cal S}_\G(\vec{k}_F- \vec{\hat{k}}, - \omega) \right]$.
 We will also define  the important  asymmetry ratio: 
 \beq
 {\cal R}_{\G}(\vec{k}_F |\vec{\hat{k}}, \omega) = {{\cal S}_\G^{a-s}(\vec{k}_F| \vec{\hat{k}}, \omega)} /{{\cal S}_\G^{s}(\vec{k}_F| \vec{\hat{k}}, \omega)}, \label{rr}
 \eeq
where  normalization factors cancel out, giving a  dimensionless function of order unity. Its magnitude  can  therefore be compared across different systems. 
 We will quote ${\cal R}_{\G}$ and ${\cal S}_\G^{s}$ below for various theoretical  models;  ${\cal S}_\G^{a-s}$ can be reconstructed  from \disp{rr}.

{\bf 3. Dynamical particle hole symmetry of the Fermi Liquid theory}.
We begin by considering ${\cal S}_\G$ for the Fermi liquid theory.     The spectral function of a Fermi liquid $ \rho^{FL}_{G}(\vec{k} ,\omega)$ is  given  in terms of a smooth background plus   a quasiparticle peak as in \disp{fl}. Near the Fermi surface we can linearize various objects in $\hat{k}$ and $ \omega$.  With  $\vec{v}_{\vec{k}_F} $  the Fermi velocity vector at $\vec{k}_F$, the quasiparticle piece is   specified by    three parameters (i) renormalization factor $Z_{\vec{k}}$,  with a linear dependence $Z_{\vec{k}}=Z_{\vec{k}_F}[1+ c_1  (  \vec{\hat{k}}.\vec{v}_{\vec{k}_F})]$,  (ii) the quasiparticle energy $E_{\vec{k}}$  vanishing linearly  at the Fermi surface $E_{\vec{k}} = \frac{m}{m^*} (  \vec{\hat{k}}.\vec{v}_{\vec{k}_F})$ with an effective mass renormalization $ \frac{m}{m^*} $  and (iii) the line width $\gamma_{\vec{k}} \propto [ E_{\vec{k}}^2 + (\pi k_B T)^2 ]$    vanishes symmetrically at the Fermi surface. Thus near the Fermi surface:
\beq
\rho^{FL}_G(\vec{k} ,\omega)\sim \rho^{(bg)}_G(\vec{k}, \omega)+ \frac{Z_{\vec{k}}}{\pi} \frac{\gamma_{\vec{k}}}{\gamma^2_{\vec{k}}+(\omega- E_{\vec{k}} )^2}.\label{fl}
\eeq
 For $\vec{k}$ close to the Fermi surface, the background part is   neglected compared to the large quasiparticle part.   Defining  the quasiparticle peak part
\beq
\Q(\vec{\hat{k}}, \omega)=  \frac{Z_{\vec{k}_F}}{ 4 \pi \cosh^2(\beta \omega/2)}
 \frac{\gamma_{\vec{k}_F}}{\gamma^2_{\vec{k}_F}+(\omega- \frac{m}{m^*} (  \vec{\hat{k}}.\vec{v}_{\vec{k}_F}) )^2}  , \label{q}
\eeq
 we write the Fermi symmetrized functions of $ (\vec{\hat{k}}, \omega)$ :
\barray
\{ {\cal S}^{s}_{G_{FL}}, {\cal R}^{}_{G_{FL}}\}  & =&   \{ \Q(\vec{\hat{k}}, \omega),   c_1  (  \vec{\hat{k}}.\vec{v}_{\vec{k}_F}) \} \label{s-fl},
\earray
where we retained  only  terms linear in $\hat{k},\omega$ beyond the quasiparticle peak term ${\cal Q}(\vec{\hat{k}}, \omega) $.
Observe that  to $O(\omega^2)$ the asymmetry ratio  ${\cal R}$ {\em is  independent of} $\omega$.  The  requirement of neglecting the background is necessary, since it is hard to make a general statement about the $(k,\omega)$ dependence of the background part.  Therefore the discussion  becomes sharp only in situations where the peak term overwhelms the background part- thus forcing us to low temperatures. The same issue also impacts the synchrotron data adversely compared to the laser ARPES data, if we interpret the former to have more substantial elastic scattering correction as argued in \refdisp{gweon}.

We make a few remarks next.
{\bf (1)} The coefficient $c_1$ vanishes in theories where the self energy is $\omega$ dependent but  $\vec{k}$ independent.  To the extent that we can experimentally identify a $\omega$ independent but $k$ dependent term as in  \disp{s-fl}, {\em one can say that  the Fermi liquid spectrum  possesses the dynamical particle hole invariance}.
{\bf  (2)}
{ The  momentum dependence of the dipole transition probability $M(\vec{k})$, if any,  can be absorbed into  $c_1$ in \disp{s-fl} by Taylor expansion.  This implies that  the expression \disp{s-fl} is valid for the ${\cal S, R}$ constructed from the ARPES intensities directly (i.e. omitting the $1/ M$ term in \disp{symm}). \em  The important  asymmetry ratio ${\cal R}$ gets rid of the overall scale factors. Therefore  its magnitude is a  meaningful quantitative  measure of the asymmetry.}
{\bf (3)}
It follows that the  frequency independence of ${\cal R}$ is also true for any theory where the Dyson self energy   $\Im m  \ \Sigma(k,\omega)$ is  {\em even}   (i.e. not necessarily  quadratic)  in  $\omega$, such as the marginal Fermi liquid \cite{mfl}  and  also various refinements of the RPA. Subleading corrections of the type $\omega \times T^2$ or $\omega^3$  in  $\Im m \ \Sigma(k,\omega)$ \cite{hodges}, as well as  intrinsic particle hole asymmetric density of states terms can lead to a non trivial 
${\cal R}$ .  However these are estimated \cite{hodges,fndisorder} to be an order of magnitude smaller than the predicted  asymmetry of the theories discussed next.

{ \bf 4. The asymmetry  ratio in  ECFL: \label{sec3}}
In the recent work on the ECFL\cite{ecfl}   $ \rho_\G(\vec{k},\omega) $, is the product of a Fermi liquid spectral function $\rho_{\GH}(\vec{k},\omega )$ and a   caparison factor
$ \left(   \left\{ 1- \frac{n}{2} \right\} \ +
 \frac{\xi_k - \omega }{\Delta(\vec{k},\omega)}+\eta(\vec{k},\omega )  \right) $, explicitly containing  a linear dependence on the energy $\omega$.  This important term redistributes the  dynamical spectral weight {\em within the lower Hubbard band}, in such a way as  to preserve the Fermi volume. In a further   approximation of the formalism, a  simplified ECFL theory emerges where we obtain  explicit analytical results.  In this version   $\eta(\vec{k},\omega )$ is negligible  and the coefficient $ \Delta$ is a constant determined by the number sum rule. In  \refdisp{gweon,anatomy}, the   simplified ECFL  was tested against  data on the High $T_c$ cuprate $Bi_2Sr_2CaCu_2 O_{8+\delta}$.  The test spans  a substantial   range   of occupied energies  $\sim 1$ eV,  with quantitative fits in the  $0.25$ eV energy range. 
  The remarkably close agreement  between data and theory over the broad range of data sets   appears to  vindicate  the form of the spectral function. The   test proposed in this work is somewhat complementary, it is over a smaller energy range $\sim 2 k_B T$,  {\em probing the   asymptotic  low energy  region centered around the   Fermi energy}.

 With the assumption of a smooth $k$ dependence of  $\eta(\vec{k}, 0)$ and  $\Delta(\vec{k},0)$ in the expression for the spectral function \cite{comment} and $p=d_0+  (1- \frac{n}{2})$, we obtain 
  \barray
& &{\cal S}^{}_{\G_{ECFL}} \sim \Q
 \times [ p +d_1 \ \vec{\hat{k}}.\vec{v}_{\vec{k}_F} +d_2 \ \omega +  \frac{(\vec{\hat{k}}.\vec{v}_{\vec{k}_F} - \omega )}{\Delta(\vec{k}_F)} ]. \label{asymm-exact}\nn
\earray
\begin{figure}
\includegraphics[width=5in]{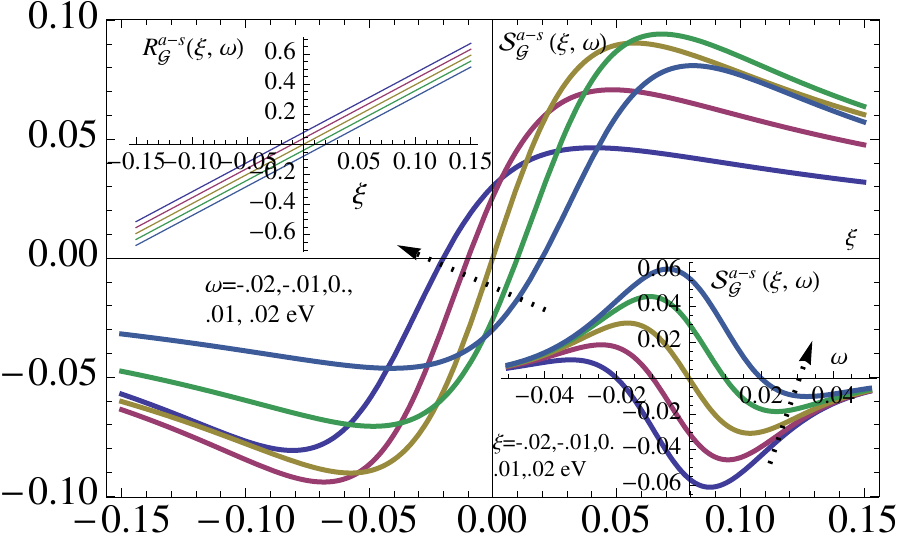}
\caption{  Top inset shows the   large predicted  asymmetry ${\cal R}_{\G_{SECFL}}^{a-s}$ versus $\xi$ in  the  small energy range of 150 meV. Similar magnitudes are found as functions of  $\omega$ at various $\xi$.
 The figure  shows  ${\cal S}_{\G_{SECFL}}^{a-s}$ from  \disp{s-hd}. 
 versus $\xi$ (main) $\omega$ (inset) in eV at various $\omega$ (main) $\xi$ (inset). Arrows indicate the direction of increasing energies.    We used $n=0.85$, $\eta=.05$ eV, $\Delta_0=.0796$ eV here. }
\label{Fig_1}
\end{figure}
Here the term $d_0$ arises from Taylor expanding $\eta(\vec{k}_F, 0)$  and also the shift of the chemical potential from the free value, $d_1$ from the momentum dependence of $Z_k$ and  this term can also absorb the momentum dependence of    $M(k)$, and  $d_2$ from the frequency dependence of $\eta(k,\omega)$.    We can thus compute the symmetric and anti-symmetric parts $ \{ {\cal S}^{s}_{\G_{ECFL}}, {\cal R}^{}_{\G_{ECFL}}\} $ as:
\barray
&&\sim  \{ p \  \Q, \frac{d_1}{p}  \ \vec{\hat{k}}.\vec{v}_{\vec{k}_F} + \frac{d_2}{p} \ \omega+   \frac{(\vec{\hat{k}}.\vec{v}_{\vec{k}_F} - \omega )}{p \ \Delta(\vec{k}_F)} \}.\label{s-ecfl}
\earray
The asymmetry ratio ${\cal R}$ therefore  has a linear $\omega$ and $\hat{k}$ dependence.
Using  the    frequency dependence as the signature, 
 one should be able to distinguish between the results of \disp{s-ecfl} and \disp{s-fl} .

  The  simplified ECFL model (SECFL)   is   described in detail in   \refdisp{anatomy}, where   we  write the spectral function near the Fermi energy $ \rho_{\G_{SECFL}}^{Peak}(\vec{k}_F+ \vec{\hat{k}},\omega ) $ as:
\beq
 \frac{1}{\pi} \frac{Z_{k}^2 \ \Gamma_k}{Z_{k}^2 \ \Gamma_k^2+(\omega-E^{FL}_k)^2} \ \frac{n^2}{ 4 \Delta_0} \ \left\{ \varepsilon_0 + \xi_k -\omega \right\}.\label{rhopeak}
\eeq
where $\varepsilon_0= \Delta_0 \ \frac{4}{n^2} (1 - \frac{n}{2})$.  Here $E_k^{FL}= Z_k \ \xi_k $, in view of the form of the  self energy $\Phi$. 
To leading order, we can set $Z_k\to Z_F$ independent of $k$, and 
$\xi_k= \vec{\hat{k}} . \vec{v}_{\vec{k}_F}  $, $E_k^{FL}= Z_F \  \vec{\hat{k}} . \vec{v}_{\vec{k}_F}$, and set
 $\Gamma_k = \eta+ \pi C_\Phi [ (\pi k_B T)^2+ (E_k^{FL})^2] $, where $\eta$ is the elastic broadening introduced in \refdisp{gweon} (distinct from $\eta(\vec{k},\omega)$). For the  model \disp{rhopeak},
 we can set $\Gamma_k\to \Gamma_{k_F}$ 
 and thus obtain the leading behavior near the Fermi energy of $ \{  {\cal S}^{s}_{\G_{SECFL}},  {\cal R}_{\G_{SECFL}} \} $ as:
 \barray
&\sim&[ (1- \frac{n}{2})  \Q(\vec{\hat{k}}, \omega),   \frac{\{ \vec{\hat{k}} . \vec{v}_{\vec{k}_F}   -\omega \}}{\varepsilon_0} ],  \label{s-hd}
 \earray
 where $\Q(\vec{\hat{k}}, \omega)$ is obtained from   \disp{q} by replacing $m/m^* \to Z_{F}$ and $\gamma_k \to \Gamma_k Z_k$. {\em Note that e.g. at $\vec{\hat{k}}=0$ and any convenient $\omega_0$,   $|{\cal R}(0,\omega_0)|= \omega_0/\varepsilon_0$,  and thus its magnitude yields  the important energy scale $\Delta_0$}. 
 We emphasize that \disp{s-ecfl} is  more generally true within the ECFL approach.    We display ${\cal S}^{a-s}$   in the  \figdisp{Fig_1}  for a model calculation based in the simplified  ECFL model with a flat density of states \refdisp{anatomy} Sec.(IV.F).  The values of the basic parameters in all figures are as follows:  $T=180$K, $\omega_c=.25$ eV, $C_\Phi=1$(eV)$^{-1}$.   
Notice the distinctive  increasing linear behavior with $\vec{\hat{k}}$  and a decreasing linear one with $ \omega$,  as in \disp{s-ecfl} and \disp{s-hd}.

 \begin{figure}[t]
\includegraphics[width=5in]{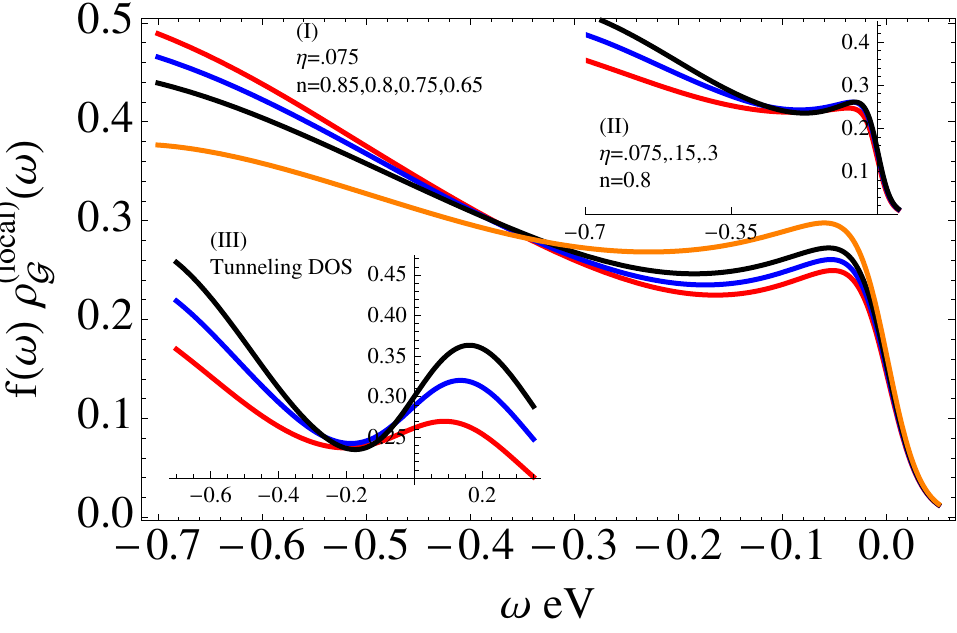}
\caption{  (I)   the predicted  AIP   spectrum  showing a shallow minimum at  $\omega\sim -.2$  eV, and a  rise as the binding energy $|\omega|$ increases. The rise is greater as    the particle density $n$ increases (orange to red). Inset (II) reveals the role of elastic scattering width  $\eta$  (black to red). Inset (III) shows the local DOS relevant to the tunneling conductance, for the same parameters as in (II) with a  remarkable rising piece  near zero bias.
}
\label{Fig_2}
\end{figure}

{\bf 5. Single particle tunneling into the extremely correlated state: \label{sec4}}
In the simplest model of tunneling in the \tJ model, the conductance is given in terms of  the local density of states  (DOS)~ $ \rho_{\G}^\l(\omega) =  \sum_{\vec{k}} \rho_{\G}(\vec{k} ;\omega)$. Its  convolution with $f_\omega$  and  $\overline{f}_ \omega$ gives  half the occupied  $\frac{n}{2}$, and the unoccupied  ($1-n$)  densities, thus  providing useful sum-rules for tunneling\cite{mohit}.  The sum rule imply asymmetry between adding particles and holes and thus a downward sloping conductance\cite{ong,hanaguri}.  Recent experiments in the overdoped regime\cite{yazdani,renner}  display  the same asymmetry, providing strong  confirmation  that \tJ model type extreme correlations are operative at high hole  doping levels as well, and  not just near half filling. More detailed information on the frequency dependence is clearly  of experimental interest. We note that the angle integrated photo emission (AIP)  technique obtains the local DOS   $\times f_\omega$, and provides a complementary view to tunneling.   \figdisp{Fig_2} presents the results from the simplified ECFL model for both the (local) DOS and  DOS $\times f_\omega$   at various densities and elastic scattering parameter $\eta$. It shows an overall decrease of the local density of states with energy. Interestingly the tunneling curve in the inset (III)  shows an upturn followed by  a rising piece near $\omega\sim0$, and the AIP curve shows a related  shallow minimum at $\omega \sim -.2$ eV.

  To understand the  unusual  result, consider integrating  the spectral function in \disp{rhopeak} over $\xi_k$.  As discussed in \refdisp{gweon,anatomy}, when the  energy is  less than $\sim 1 $eV,  the quasiparticles become sharp and  this integral  can be estimated by replacing  the Fermi liquid Lorentzian by  $ \delta(   \vec{\hat{k}} . \vec{v}_{\vec{k}_F} - \frac{m^*}{m} \omega)$.  This  yields the quasi particle peak contribution:
\beq
\lim_{\omega \leq \varepsilon_0} \rho^\l_{\G,{Peak}}(\omega) \sim   \mbox{(const)} \left\{ \varepsilon_0 + (\frac{m^*}{m} -1) \times \omega \right\}. \label{counter-intuitive}
\eeq
 Since $m \leq m^* $,  it follows that the slope is positive and hence the rising conductance! In the  general version of ECFL, different parts of the Fermi surface contribute according to the weight of $1/\Delta(\vec{k}_F)$.  We  expect   the resulting average   to be less favorable to a rising term than in the simplified ECFL  model.

{\bf 6. Other theories:} Casey and Anderson \refdisp{casey}(CA)    provide a spectral function   that may be Taylor expanded at finite $T$ and  low  enough energies as follows.  With $q=1-\frac{1}{4}n^2$ depending on the filling $n$, and $\Gamma_{\hat{k}} = A (k_B T)+ C v^2_{k_F} \hat{k}^2$, their expressions yield:
\barray
\{ {\cal S}^{s}_{CA},{\cal R}^{a-s}_{CA}  \}& =&  \{{\cal Q}' , \cot ( q  {\pi}/{2} ) 
\frac{ (v_F \hat{k} - \omega)}{\Gamma_0  } \}.
\earray
with ${\cal Q}' = \mbox{const}\times \frac{ \sin ( q  {\pi}/{2}  )  }{4 \pi  \cosh^2(\beta \omega/2)   }/\left[ \Gamma_0^2 + (\omega- v_F \hat{k})^2\right]^{q/2}$.
Therefore   this work also implies a non trivial ${\cal R}$ with a  linear $\omega, \hat{k}$ dependence,   similar in form  to that in ECFL,  although  with a non Lorentzian peak factor replacing the $\Q$ factor in \disp{s-ecfl}.
 It is seen  that the asymmetry of  this theory as well as that of the ECFL theory vanishes continuously at  {\em low particle density}  $n \to 0$.  An important  characteristic energy $\Delta^*(x,T)$,  say  the inverse of the slope of the linear in $\omega$ term in ${\cal R}$ contains much physics.  In the CA theory  $\Delta^*(x,T  )\propto \Gamma_0$      vanishes   {\em  at all densities} $x$ as $T\to 0$,  thereby defining a line of quantum critical points.  On the other hand  in the  ECFL calculations,  the energy $\Delta^*(x,T\to 0)$ is   non zero but  much smaller than the (bare) Fermi energy.  However it could vanish at a specific filling $x_c$: as $\Delta^*(x_c, T\to 0)\to 0$, thereby locating  an isolated quantum critical point. 
  
 Other contemporary theories have a different prediction from the ECFL and CA.  The popular marginal Fermi liquid model\cite{mfl}   for the spectral function has a  Dyson self energy that  is {\em symmetric} under the transformation \disp{dpht}.
 Therefore it  leads to an $\omega$ independent   asymmetry ratio   at small energies, as  in  the usual Fermi liquid\cite{fndisorder}. A similar $\omega$ independent  ${\cal R}$ occurs  for the RPA and its many variants emphasizing fluctuation contributions.

\begin{figure}
\includegraphics[width=2.75in]{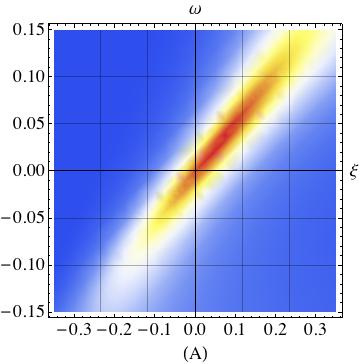}
\includegraphics[width=2.75in]{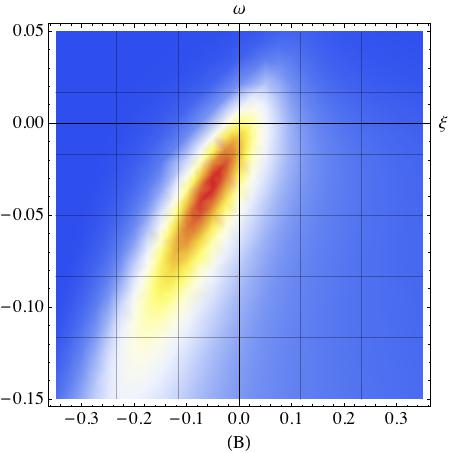}\\
\includegraphics[width=2.75in]{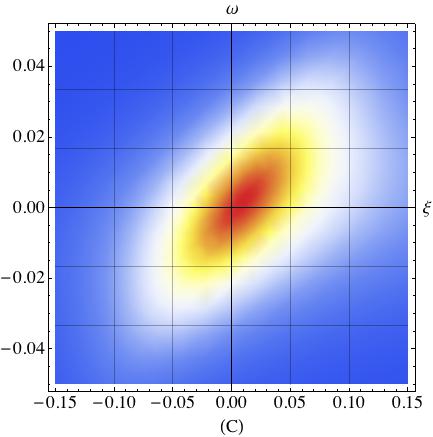}
\includegraphics[width=2.75in]{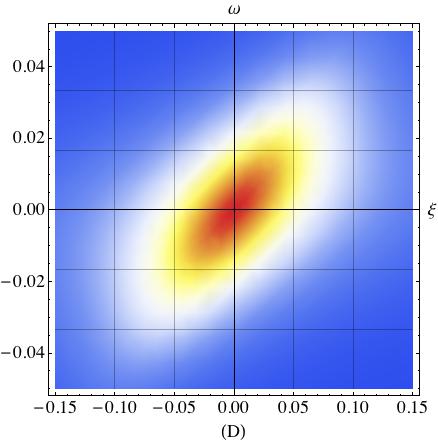}\\
\includegraphics[width=2.75in]{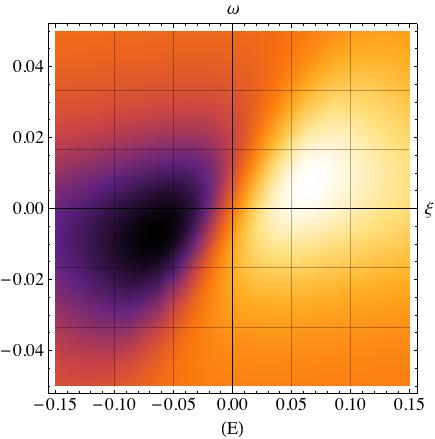}
\includegraphics[width=2.75in]{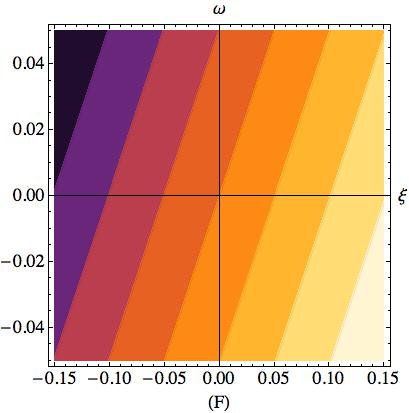}
\caption{ Symmetry extraction illustrated for the  simplified  ECFL model.   Here $n=0.85$,  $\eta=.05$ and $\Delta_0=.0796$  with $\omega$ (ordinate)  and $\xi$  (abscissa) in eV.   (A)  shows the spectral function $\rho_{\G}$, 
(B)  $\rho_{\G} f_\omega $,   (C) $\rho_{\G} f_\omega \overline{f}_\omega $    (D) the  symmetrized object ${\cal S}_{\G}^{s}$   (E) the antisymmetrized object
   ${\cal S}_{\G}^{a-s}$  showing a peak and a trough,
   and (F)  the asymmetry ratio ${\cal R_{\G}}$ from \disp{rr}.}
\label{Fig_3}
\end{figure}

{\bf 7. Conclusions \label{sec5}} The program of  extraction of the asymmetry ratio from the ``ideal''  spectral weight   is  summarized in \figdisp{Fig_3}.   A window of size $\sim 2 k_B T$ in  $\omega$ and  $v_F \hat{k}$    is highlighted in this construction.  It is   proposed  that a careful examination of the ARPES intensity along these lines would determine  the  existence of  dynamical particle hole asymmetry.   This asymmetry also relates to the difference in velocities (and amplitudes) of quasi particles and quasi holes, of the type that    are  invoked  in explaining  the peculiar sign of the  Hall effect in the mixed state\cite{hall}. We thus expect it to be important in Hall and analogous   transport contexts such as thermopower.    This search is complementary, as well as  a pre-requisite,  to  the  detailed     characterization of  the   symmetric part ${\cal S}^{s}$.  Specifically I   propose that  the search for  a non trivial (i.e. $\omega$  linear)  asymmetry ratio ${\cal R}$ is important  for identifying the  correct underlying theoretical  description of the cuprates.  
  
 In order to implement the transformation \disp{dpht} on the experimental data, we need  a high resolution in  frequency as well as   momentum.  Since the bare Fermi velocities are  high  
 ~ $ \hbar v_F \sim 5$ eV \AA, the momentum resolution becomes critical.   An error  
 ~ $\Delta \xi  \sim 15-20$ meV  can  lead to quite incorrect conclusions. Thus in order to draw unambiguous conclusions  we require   $\Delta k \sim .001 ( \AA)^{-1}$,  i.e.  $\Delta \xi \sim 5$ meV or better,   thereby  posing an interesting challenge to the experimental ARPES community.

{\bf Acknowledgements}
  This work was supported by DOE under Grant No. FG02-06ER46319. I  thank  A. Chubukov, A. Georges, G. H. Gweon,   D. Huse, H R. Krishnamurthy  and T. V. Ramakrishnan for stimulating comments.


\begin{thebibliography}{36}
\bibitem{fndisorder}    A  small asymmetry is inevitably induced by  density of states variation leading to an estimated   asymmetry ratio ${\cal R}$  (\disp{rr}) that is an order of magnitude smaller than that found in the ECFL and CA theories.   Also the unavoidable disorder contributes, and thus one may operationally refer to  cases as essentially invariant  if    $|{\cal R|}  \leq 1 \%$ in an  energy window  of $\pm 25 \ meV$  around $\chem$ .
\bibitem{comment}
 The energy variables  $\Delta(\vec{k},\omega) $ and  $\eta(\vec{k},\omega)$  are defined in Eqs.~(21,22,23) of \refdisp{ecfl}.
\end{thebibliography}
\end{document}